\documentclass[]{WileyChemistry-template}

\usepackage[version=3]{mhchem} 
\usepackage{braket}
\usepackage{color}
\usepackage[dvipsnames]{xcolor}
\usepackage{textgreek}
\usepackage{amsmath,amssymb}
\usepackage{amsfonts}
\usepackage{mathtools}
\usepackage{bm}

\title{\raggedright Local Chiral Optical Responses in Phthalocyanine Molecular Assemblies Revealed by Photoinduced Force Microscopy}

\author{
\begin{minipage}{\textwidth}
	Masayoshi Fujii,\textsuperscript{[a]} 
    Mamoru Tamura,*\textsuperscript{[a,b,c,d]}
    Hidemasa Yamane,\textsuperscript{[e]}
    Hajime Ishihara,*\textsuperscript{[a,d,f,g]}
\end{minipage}
}

\newcommand{\affiliation}{
\begin{itemize}

\item[{[a]}] M. Fujii, Dr. M. Tamura, Prof. Dr. H. Ishihara\\
Department of Materials Engineering Science, The University of Osaka, 1-3 Machikaneyama-cho, Toyonaka, Osaka 560-8531, Japan

\item[{[b]}] Dr. M. Tamura*\\
Department of Physics and Astronomy, School of Science, Kwansei Gakuin University, 1 Gakuen Uegahara, Sanda, Hyogo 669-1330, Japan\\
E-mail: mamoru.tamura@kwansei.ac.jp

\item[{[c]}] Dr. M. Tamura\\
Research Institute for Light-induced Acceleration System (RILACS), Osaka Metropolitan University, 1-2 Gakuencho, Nakaku, Sakai, Osaka 599-8570, Japan

\item[{[d]}] Dr. M. Tamura, Prof. Dr. H. Ishihara*\\
Research Organization of Science and Technology, Ritsumeikan University, 1-1-1 Nojihigashi, Kusatsu, Shiga, 525-8577, Japan\\
E-mail: ishi-h@fc.ritsumei.ac.jp

\item[{[e]}] Dr. H. Yamane\\
Osaka Research Institute of Industrial Science and Technology, 2-7-1 Ayumino, Izumi, Osaka 594-1157, Japan

\item[{[f]}] Prof. Dr. H. Ishihara\\
\mbox{R\textsuperscript{3}} Institute for Newly-Emerging Science Design, The University of Osaka, Toyonaka, Osaka 560-8531, Japan

\item[{[g]}] Prof. Dr. H. Ishihara\\
Ritsumeikan Semiconductor Application research center (RISA), Ritsumeikan University, Kusatsu, Shiga 525-8577, Japan

\end{itemize}
}

\renewcommand{\abstract}{
Photoinduced force microscopy (PiFM) provides a force-based probe of nanoscale polarization-dependent molecular excitations.
Here, we theoretically investigate PiFM images of zinc phthalocyanine assemblies using the discrete dipole approximation with nonlocal molecular susceptibilities.
Under linearly polarized illumination, intermolecular dipole coupling splits the molecular resonance into bonding and antibonding modes.
These modes produce distinct force distributions: bonding modes enhance signals at molecular termini, whereas antibonding modes localize responses in intermolecular regions, reflecting collective intermolecular polarization modes.
Under circularly polarized illumination, intermolecular coupling and anisotropic molecular packing generate spatially varying local circular dichroism signals with enhanced asymmetric force factors.
These results establish PiFM as a real-space probe of collective polarization modes and coupling-induced local chiral optical responses.
}

\newcommand{\keywords}{
	Phthalocyanines \textbullet\ 
	Scanning probe microscopy \textbullet\ 
	Circular dichroism \textbullet\ 
	Cooperative effects \textbullet\ 
	Through-space interactions
}

\begin{document}

\twocolumn[\vspace{-1.5cm}\maketitle\vspace{-1cm}
	\textit{\dedication}\vspace{0.4cm}]
\small{\begin{shaded}
		\noindent\abstract
	\end{shaded}
}

\begin{figure} [!b]
\begin{minipage}[t]{\columnwidth}{\rule{\columnwidth}{1pt}\footnotesize{\textsf{\affiliation}}}\end{minipage}
\end{figure}




\section*{Introduction}
\label{introduction}

Nanoscale optical imaging beyond the diffraction limit is important for understanding intermolecular coupling and polarization-dependent optical responses in molecular systems.\cite{Verma2017, Lee2018, Sifat2022}
Luminescence-based nanospectroscopic techniques such as tip-enhanced photoluminescence (TEPL) and scanning tunneling microscope-induced luminescence (STML) have enabled high-resolution imaging of excited states and intermolecular coupling at the single-molecule level.\cite{Imada2016, Zhang2016}
These techniques are particularly powerful for emissive molecular systems.
For non-emissive systems, alternative approaches are required.
Photoinduced force microscopy (PiFM) detects optical responses through nanoscale forces induced by light absorption\cite{Rajapaksa2010, Yamanishi2021, Yamamoto2024}.
PiFM combines optical excitation with force detection using atomic force microscopy (AFM) \cite{Sifat2022}.
The formation of a picocavity between a metallic probe tip and the substrate enables sub-nanometer spatial resolution and high sensitivity.

Recent studies have advanced the theoretical understanding of PiFM responses at the single-molecule level,
showing that electronic transitions can be resolved through local force signals.
Under circularly polarized illumination, polarization-dependent optical responses such as circular dichroism have also been predicted and experimentally investigated using PiFM~\cite{Yamane2023,Yamanishi2023ChiroForce,Yamane2024SuperchiralPiFM}.
These studies demonstrate that PiFM can resolve polarization-dependent near-field responses in molecular systems.
In molecular assemblies, intermolecular dipole coupling and structural chirality are expected to generate collective optical responses that are absent in isolated molecules. 
However, how intermolecular coupling and structural chirality appear in PiFM images remains largely unexplored.

For example, STML studies have reported energy splitting and coupling-dependent emission 
modulation in molecular dimers and coupled molecular systems\cite{Zhang2016},
but such effects have not been extensively explored through non-luminescence-based techniques. 
In this study, we demonstrate that intermolecular dipole coupling produces characteristic spatial contrast in PiFM images of phthalocyanine molecules under linearly polarized illumination.
Furthermore, we show that circularly polarized illumination induces geometry-dependent chiral responses in PiFM images of phthalocyanine molecular assemblies.
These results clarify how intermolecular coupling, molecular geometry, and polarization conditions govern PiFM contrast at the nanoscale.

\section*{Results and Discussion}
\label{results_discussion}


\begin{figure*}
\begin{center}
\includegraphics[width=16cm]{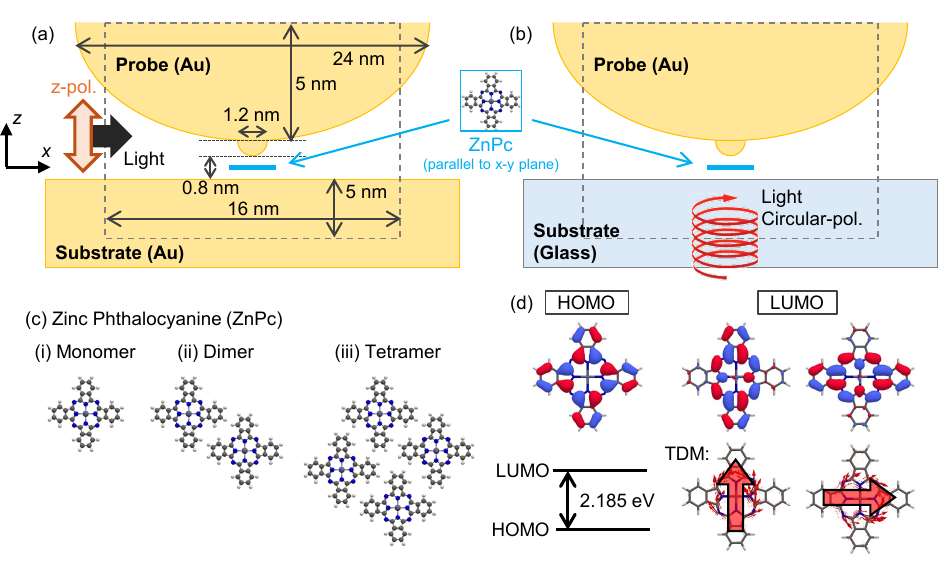}
\caption{
Schematic illustration of the simulation models and molecular structures used for the PiFM calculations. 
(a) Linearly ($z$-) polarized excitation with an Au tip and Au substrate. 
(b) Circularly polarized excitation with an Au tip and glass substrate. 
(c) Molecular configurations of ZnPc aggregates.
(d) HOMO and LUMO orbitals of ZnPc and the corresponding in-plane transition dipole moment.
}
\label{fig:model}
\end{center}
\end{figure*} 

Figure 1 illustrates the computational models used in this study to simulate PiFM images under (a) linearly polarized and (b) circularly polarized illumination. The geometries of the probe tip and substrate were identical in both models.
The probe tip and substrate were modeled within a finite computational domain with a lateral size of 10 nm $\times$  10 nm, resulting in lateral truncation of both structures at the simulation boundary. The probe tip was represented by a semi-ellipsoid with a diameter of 12 nm in the $x$ and $y$ directions and a semi-axis length of 4 nm along the $z$ direction, with a hemispherical protrusion of 1 nm diameter at the apex. The substrate was modeled as a rectangular prism with a height of 2 nm.
The tip apex and substrate surface were separated by a gap distance of 0.8 nm, and single or multiple ZnPc molecules lying parallel to the x-y plane were positioned at the center of the gap.
Although the molecules were treated as suspended in vacuum in the present computational model, thin insulating layers such as NaCl are typically introduced experimentally to suppress wavefunction hybridization between the molecules and metallic substrate. Such insulating layers were not included in the present simulations.

In Fig. 1(a), both the probe tip and substrate were assumed to be gold and irradiated with a linearly ($z$-) polarized plane wave propagating along the +x direction. The nanogap formed between the probe and substrate generates a highly localized and enhanced electric field in the visible wavelength range.
In Fig. 1(b), the probe was modeled as gold while the substrate was glass. Left- and right-circularly polarized plane waves propagating along the +z direction were used for excitation.
Figure 1(c) shows the molecular configurations of ZnPc considered in this study: a monomer, dimer, and tetramer.
In all cases, the ZnPc molecular planes were oriented parallel to the $xy$-plane.
Figure 1(d) illustrates the HOMO and LUMO molecular orbitals of ZnPc, which play dominant roles in the optical response near the resonance energy. The HOMO--LUMO transition energy obtained from the DFT calculations was 2.185 eV. The spatial distribution of the transition dipole moment density calculated from the molecular orbitals is also shown. As indicated by the large red arrows, the resulting overall transition dipole moment is predominantly oriented along the x or y direction within the molecular plane.

During tip scanning over the ZnPcs, the optical response of the nanogap strongly depends on the relative tip–substrate position because the finite simulation domain represents only a limited portion of the metallic structure. 
To avoid this artificial effect, the tip-scanning process was reproduced by changing the relative tip–substrate position through translation of the ZnPcs opposite to the tip motion.


Figure 2 shows the optical force spectra and PiFM images of ZnPc monomers (M) and dimers (D).
In Fig. 2(a), the tip positions used for calculating the spectra are indicated in the figure legend (in units of \AA), and the vertical dashed lines indicate the excitation energies used for the PiFM image calculations.
The gray dashed line M(0,0) obtained when the tip was positioned at the center of the monomer does not exhibit a peak-like profile because the coupling between the tip field and the molecular transition dipole is canceled by symmetry, as discussed below.
Although an optical force is still detected, it originates from the interaction between the tip and the substrate.

The green line M(6,0) for the tip positioned near the edge of the monomer, exhibits a resonant attractive force at 2.133~eV.
The resonance energy is lower than the intrinsic HOMO--LUMO transition energy of the isolated molecule, which is attributed to a redshift arising from the interaction of the molecule with the metallic environment, including both the tip and the substrate.
For the dimer, the tip positioned at the terminus of the dimer (red line D(12,$-4$)) exhibits a resonance peak at 2.126~eV,
whereas the tip positioned between the molecules (blue line D(0,0)) exhibits a resonance peak at 2.142~eV.
These resonances are assigned to bonding and antibonding polarization-coupling modes, respectively.

The PiFM images calculated at these characteristic excitation energies are shown in Figs.~2(b)--(d), where the red lines show the ZnPc molecular frameworks.
In the monomer image in Fig.~2(b), the signal is weak at the molecular center and forms a ring-like pattern around the molecular framework.
This behavior reflects the in-plane transition dipole moment of ZnPc.
In the present optical configuration, the near field at the tip apex is radially distributed within the molecular plane.
When the tip is positioned directly above the molecular center, the inner product between the transition dipole moment and the local optical field cancels due to symmetry, resulting in no molecular resonant contribution to the optical force.
In contrast, when the tip is displaced from the center, a non-zero inner product arises between the radial electric field and the dipole moment, yielding a finite signal.
As a result, the molecular resonant contribution is suppressed at the center and enhanced around the molecular periphery, producing the ring-like PiFM pattern shown in Fig.~2(b).

In the dimer, the PiFM image changes significantly depending on the excitation energy.
At 2.126~eV in Fig.~2(c), the signal becomes strongest when the tip is positioned at the termini of the dimer, consistent with the spatial distribution of the bonding mode.
In contrast, at 2.142~eV in Fig.~2(d), the signal is strongest when the tip is positioned between the molecules, reflecting the antibonding character of the excitation mode.
Similar spatial contrasts associated with bonding and antibonding excitation modes have also been observed in STML studies of coupled molecular systems,\cite{Zhang2016} showing qualitative agreement with the present PiFM calculations.

\begin{figure}
\begin{center}
  \includegraphics[width=85mm]{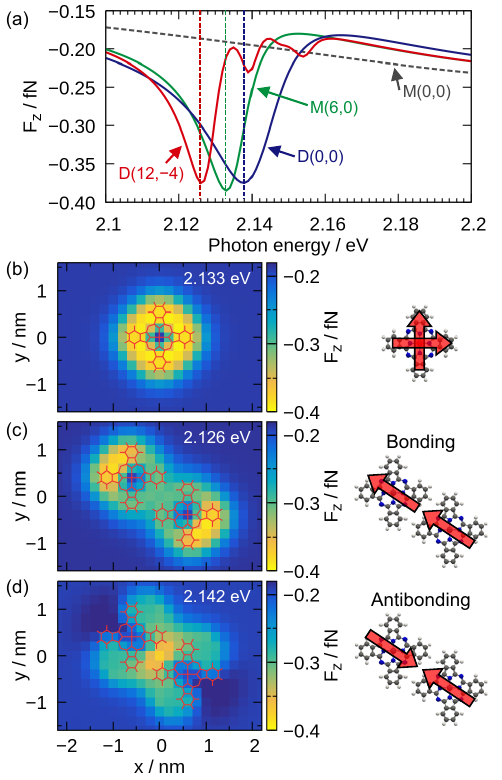}
  \caption{(a) Optical force spectra for ZnPc monomers (M) and dimers (D). (b)--(d) Calculated PiFM images: (b) monomer at 2.133~eV, (c) dimer at 2.126~eV, and (d) dimer at 2.142~eV. Arrows in the schematics indicate the induced polarizations.}
  \label{fig:LinMD}
\end{center}
\end{figure} 


Furthermore, to investigate how the spectra and PiFM images depend on the molecular arrangement of ZnPc, Fig.~3 presents results for a 2$\times$2 ZnPc array (tetramer) analyzed in the same manner as in Fig.~2.
Because several polarization modes appear in the tetramer, the schematic illustrations in Figs.~3(b)--(d) show the modes corresponding to the tip positions used in the respective spectral calculations.
The results reveal clear resonance shifts and spatial variations in the PiFM images arising from intermolecular bonding and antibonding polarization coupling.

As indicated by the red line in Fig.~3(a), when the tip is positioned at the terminus of the tetramer, a resonance appears at 2.125~eV.
Correspondingly, the PiFM image in Fig.~3(b) exhibits strong signals at the molecular termini, reflecting the bonding mode.
As indicated by the green line in Fig.~3(a), when the tip is positioned between neighboring molecules in the tetramer, a resonance appears at 2.138~eV.
The PiFM image in Fig.~3(c) correspondingly exhibits strong signals in the intermolecular regions, consistent with the spatial localization of the antibonding polarization mode.
At 2.142~eV in Fig.~3(d), a finite signal is distributed also around the center of the tetramer, suggesting a contribution from an antibonding polarization mode with diagonal components.

As demonstrated by the results in Figs.~2 and 3, PiFM can directly visualize force variations associated with intermolecular polarization coupling, providing spatially resolved insight into optical interactions and collective excitation modes within molecular assemblies.

\begin{figure}
\begin{center}
  \includegraphics[width=85mm]{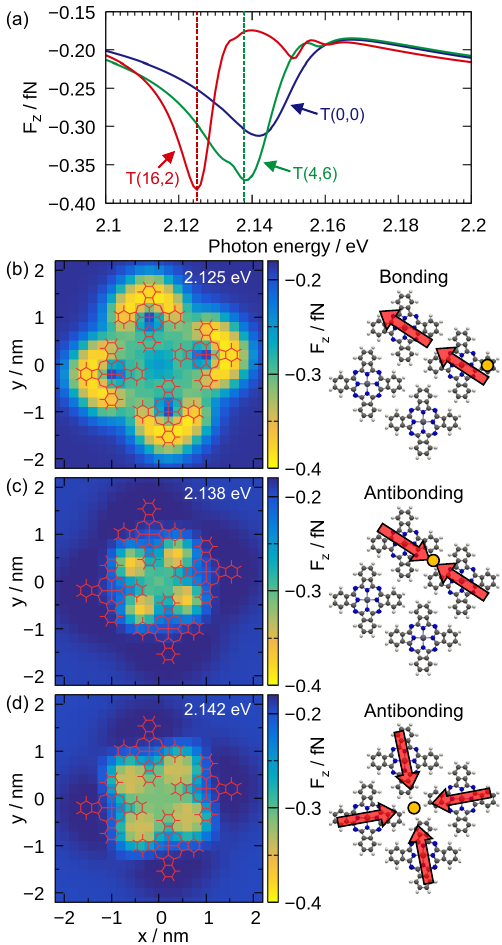}
  \caption{
(a) Optical force spectra for a ZnPc tetramer (T).
(b)--(d) Calculated PiFM images of the tetramer at (b) 2.125~eV, (c) 2.138~eV, and (d) 2.142~eV.
Arrows in the schematics indicate the induced polarizations.
}
  \label{fig:LinT}
\end{center}
\end{figure}


Next, we investigate local circular dichroism (CD) mapping using PiFM and examine how intermolecular polarization coupling influences the CD response.
Although previous studies mainly discussed monomer systems, here we analyze both monomer and coupled ZnPc molecules placed on a glass substrate under illumination with circularly polarized light incident from below.
The computational model is shown in Fig. 1(b).
As a measure of the CD response, we evaluate the following asymmetric $g$-factor:
\begin{equation}
g_f = 2 \frac{F_z^+ - F_z^-}{F_z^+ + F_z^-}
\end{equation}
where the superscripts $+$ and $-$ of $F_z^\pm$ denote the sign of the spin angular momentum of the circularly polarized light.
The light-induced force maps $F_z^+$ and $F_z^-$ used to evaluate the $g_f$ distributions in Figs.~4 and 5 are shown in Figs.~S1 and S2, respectively, in the Supporting Information.

Figure~4 shows the spectra of the asymmetric factor $g_f$ and the corresponding spatial distributions obtained by scanning the tip.
First, focusing on the monomer case at 2.164~eV in Fig.~4(b), the resulting $g_f$ map exhibits a sign-alternating distribution around the molecule.
At first glance, this behavior may appear counterintuitive because the PiFM image of a ZnPc monomer has an almost ring-like shape.
However, the apparent ring-like PiFM contrast does not necessarily imply a continuously isotropic optical response.

Okamoto discussed the local optical activity of two-dimensional structures in relation to their local symmetry. \cite{Okamoto2019}
In this argument, local CD is inactive on symmetry axes, where the responses to left- and right-circularly polarized light are equivalent.
In contrast, at positions away from the symmetry axes, the local environment can become effectively chiral, and local CD signals are allowed.
Thus, squares and regular polygons can exhibit local CD signals at positions away from their symmetry axes, whereas an ideal circular disk is locally CD-inactive because it is isotropic in the two-dimensional plane and every point is accompanied by a symmetry axis. \cite{Okamoto2019}

The present ZnPc monomer should therefore not be regarded as an ideal circular disk, even though its PiFM image appears ring-like.
As illustrated schematically in Fig.~4, the relevant optical response is governed by two orthogonal transition dipoles within the molecular plane.
Consequently, the local symmetry of the optical response is discrete rather than continuously isotropic: the transition-dipole directions and the diagonal directions act as effective symmetry axes in the molecular plane.
Local CD signals are therefore suppressed on these symmetry axes, whereas positive and negative signals appear in the regions between them.
This gives rise to the sign-alternating square-like local response observed in the calculated $g_f$ map in Fig.~4(b), rather than the no-CD response expected for an ideal circular disk.

For the dimer, the $g_f$ map at 2.156~eV in Fig.~4(c) shows positive and negative domains arranged along the molecular pair.
This distribution indicates that intermolecular polarization coupling makes the effective optical response more anisotropic in the molecular plane.
As illustrated in the dimer schematic in Fig.~4, the coupled dimer behaves not simply as two ring-like molecules, but as an elongated anisotropic system with distinct longitudinal and transverse polarization responses.
At 2.156~eV, the longitudinal bonding polarization along the molecular-pair axis is strongly induced, producing a rectangular-like local CD response with spatially separated positive and negative domains.

At 2.167~eV in Fig.~4(d), a similar sign-alternating distribution is observed, but the response is more concentrated in the intermolecular region and differs from that at 2.156~eV.
This difference reflects the stronger excitation of a transverse polarization mode along the short-axis direction of the dimer.
Although the longitudinal component still retains bonding character, it is less strongly excited at this energy.
Thus, the different spatial distribution of $g_f$ mainly reflects the change in the dominant polarization component, while the underlying symmetry of the dimer remains unchanged.
The slight blueshift from the monomer resonance can be attributed to weak antibonding coupling between the transverse polarization modes of neighboring molecules.

At the higher excitation energy of 2.185~eV in Fig.~4(e), the $g_f$ distribution becomes more complex.
This behavior is attributed to the contribution of an antibonding polarization mode, together with the selective sampling of coupled
molecular polarizations by the highly localized tip field.

\begin{figure}
\begin{center}
  \includegraphics[width=85mm]{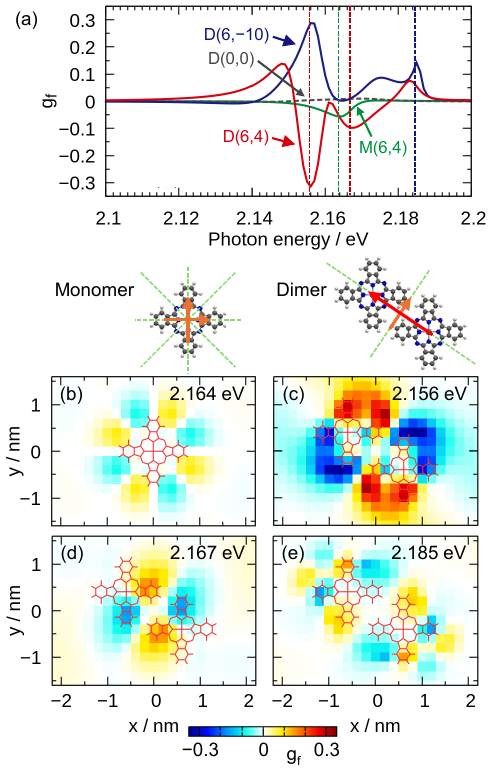}
  \caption{(a) $g_f$ spectra for ZnPc monomers (M) and dimers (D).
(b)--(e) Calculated $g_f$ maps obtained by scanning the tip at fixed excitation energies: (b)  ZnPc monomer at 2.164 eV, and (c)--(e) ZnPc dimers at 2.156, 2.167 and 2.185 eV, respectively.
The schematics illustrate the effective in-plane polarization symmetry of the monomer and dimer; dashed lines and arrows indicate local symmetry axes and representative polarization directions, respectively.
}
  \label{fig:CirMD}
\end{center}
\end{figure}


Fig.~5 shows the results for the 2$\times$2 ZnPc array (tetramer).
Because the $g_f$ spectra depend sensitively on the molecular arrangement, multiple photon energies exhibiting large $g_f$ values appear in Fig.~5(a).
Here, we focus on representative spatial distributions that are characteristic of this tetramer packing.

At 2.140 and 2.154~eV in Figs.~5(b) and 5(c), respectively, the $g_f$ maps exhibit similar sign-alternating distributions around the tetramer.
These patterns are associated with bonding polarization modes of the tetramer, indicating that the 2$\times$2 ZnPc array behaves as a coupled aggregate through intermolecular polarization coupling.
In these bonding modes, positive and negative $g_f$ signals appear around the molecular assembly, resembling the local sign reversal observed for the monomer in Fig.~4(b).

At 2.176~eV in Fig.~5(d), the $g_f$ map exhibits a more complex spatial distribution, similar to the higher-energy dimer response shown in Fig.~4(e).
This pattern is associated with an antibonding polarization mode of the tetramer.
Interestingly, a finite $g_f$ signal is also observed near the center of the tetramer.
This response can be understood as arising from the anisotropic packing geometry and collective polarization of the tetramer, which produce a locally asymmetric optical response within the molecular assembly.

These results demonstrate that intermolecular polarization coupling gives rise to characteristic local CD distributions associated with bonding and antibonding polarization modes, which can be spatially resolved by PiFM.

\begin{figure}
\begin{center}
  \includegraphics[width=85mm]{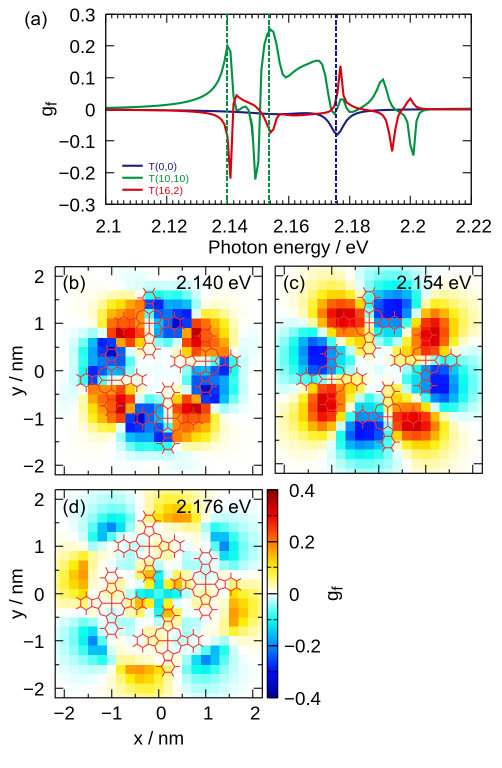}
\caption{(a) $g_f$ spectra for a ZnPc tetramer (T).
(b)--(d) Calculated $g_f$ maps obtained by scanning the tip at fixed excitation energies: (b) 2.140 eV, (c) 2.154 eV, and (d) 2.176 eV.
The colorbar of (d) is common to (b)--(d).
}
  \label{fig:CirT}
\end{center}
\end{figure}

\section*{Conclusion}
\label{conclusion}

In summary, we theoretically investigated optical force spectra, PiFM images, and local circular dichroism responses in ZnPc molecular assemblies using the discrete dipole approximation combined with nonlocal molecular susceptibilities.
For molecular dimers and tetramers under linearly polarized illumination, bonding and antibonding polarization-coupling modes produced characteristic spatial distributions in both the optical force spectra and PiFM images.
In particular, bonding modes generated enhanced signals at the molecular termini, whereas antibonding modes produced localized responses in intermolecular regions and additional spatial variations depending on the molecular arrangement.

Under circularly polarized illumination, intermolecular coupling and anisotropic molecular packing generated characteristic local CD distributions accompanied by enhanced $g_f$ responses.
The coupled molecular assemblies exhibited spatially varying chiral contrasts, indicating that collective polarization coupling and molecular arrangement strongly modify local chiral optical responses.

These results demonstrate that PiFM can spatially resolve collective polarization modes and coupling-induced local chiral optical responses in molecular assemblies, providing insight into nanoscale intermolecular optical interactions.

\section*{Computational Methods}
\label{experimental}


To analyze the optical response of probe metal tip and the molecule, we employed the discrete dipole approximation (DDA) method \cite{Purcell1973DDA, Yurkin2007DDAReview, Chaumet2022DDAReview}.
 Usually, it is described within the framework of the local response theory, which assumes that the polarization at a position $\mathbf{r}$ depends only on the electric field $\mathbf{E}(\mathbf{r},\omega)$ at the same point. However, at the nanometer scale, when the transition between quantum states in the object is important for the optical response for example in the semiconductor quantum dot and the molecules considered here, it could be incorporated into the nonlocal susceptibility based on the nonlocal response theory.\cite{Tomoshige2025TEPL, Tomoshige2025Nanophotonics, Ikagawa2026} In the DDA, the simulation region was discretized into many cells. At the cell indexed i and located at $\mathbf{r}_i$, the response electric field $\mathbf{E}_i$ and induced polarization $\mathbf{P}_i$ can be expressed by,

\begin{equation}
  \mathbf{E}_i(\omega) = \mathbf{E}_{\text{inc}}(\mathbf{r}_i, \omega) 
  + \sum_j \mathbf{G} (\mathbf{r}_i - \mathbf{r}_j, \omega) \mathbf{P}_j(\omega) V,
  \tag{2}
\end{equation}
 
\begin{equation}
  \mathbf{P}_i(\omega) =
  \begin{dcases}
    \varepsilon_0 \sum_{j \in \text{mol}} \chi_\mathrm{mol}(\mathbf{r}_i, \mathbf{r}_j, \omega) \mathbf{E}_j(\omega), & (i \in \text{mol}) \\
    \varepsilon_0 \chi (\mathbf{r}_i, \omega)\mathbf{E}_i(\omega), & (\text{otherwise})
  \end{dcases}
  \tag{3}
\end{equation}

\noindent
where $\mathbf{E}_\mathrm{inc}(\mathbf{r},\omega)$ is the incident electric field,
$\mathbf{G}(\mathbf{r};\omega)$ is the Green's function of electromagnetic field,
and $V$ is the volume of discretized cell.
Note that when treating the scattered field inside a molecule in Eq. (2), the effect of the longitudinal field is already embedded in the nonlocal susceptibility through the quantum-chemical calculations.
Therefore, for the cells corresponding to the region occupied by the molecule, denoted by indices $i,j\in \mathrm{mol}$,
we subtract from the right-hand side of Eq. (2) the longitudinal-field component
$\sum_{j \in \mathrm{mol}} \mathbf{G}(\mathbf{r}_i,\mathbf{r}_j, 0)\mathbf{P}_j(\omega)$
which is obtained by evaluating the Green’s function at zero frequency for $\mathbf{E}_{i \in \mathrm{mol}}(\omega)$.
The induced polarization at the cell assigned for molecules is expressed with the nonlocal susceptibility of molecule, while in the other region, the susceptibility is expressed with the conventional local form, as follows,

\begin{equation}
  \chi_{\text{mol}}(\mathbf{r}_i, \mathbf{r}_j, \omega) =
  \frac{1}{\varepsilon_0 V}
  \sum_n \sum_{m}
  \frac{\mathbf{d}^*_{n,m}(\mathbf{r}_i) \otimes \mathbf{d}_{n,m}(\mathbf{r}_j)}
  {\hbar \omega_{n,m} - \hbar \omega - i \gamma_{\text{mol}}},
  \tag{4}
\end{equation}

\begin{equation}
  \chi (\mathbf{r}, \omega) =  \varepsilon(\mathbf{r}, \omega) - \varepsilon_{\mathrm{bg}},
  \tag{5}
\end{equation}

\noindent
where, $\hbar \omega_{n,m}$ is the transition energy,
$\hbar \gamma_\mathrm{mol}$ is the damping constant,
$\mathbf{d}_{n,m}$ is the transition dipole moment,
$\varepsilon$ is the dielectric constant for all regions other than the molecule,
and $\varepsilon_\mathrm{bg}$ is the background dielectric constant.
In the metal region, the dielectric constant is described by the Drude–Critical Point (DCP) model, whose parameters are fitted to experimentally measured optical constants \cite{Vial2008DCP,Johnson1972NobleMetals}.
A constant refractive index of 1.5 is assumed for the glass substrate.
The distribution of the transition dipole moment is determined using the local dipole moment $\delta \boldsymbol{\mu}_{nm}^i$ in each cell at position $\mathbf{r}_i$, expressed as\cite{Zhang2020, Lee2023, Tomoshige2025TEPL, Tomoshige2025Nanophotonics, Ikagawa2026},

\begin{equation}
  \delta \boldsymbol{\mu}_{nm}^i  = \frac{ie}{\omega_{nm}} \int_{V_i} d\mathbf{r}_\mathrm{s}\ \mathbf{j}_{nm}(\mathbf{r}_\mathrm{s})
  \tag{6}
\end{equation}

\begin{equation}
  \mathbf{j}_{nm} = -\frac{i \hbar}{2 m_\mathrm{e}}
    \left[ \phi_m^* (\bm r) \nabla \phi_n(\bm r) - \phi_n^* (\bm r) \nabla \phi_m (\bm r) \right]
  \tag{7}
\end{equation}

\noindent
where, the $\phi_n$ is the molecular orbital of level $n$ obtained from the quantum chemistry software GAMESS(US)\cite{Schmidt1993GAMESS}.

From the calculated electric field and polarization, the photoinduced force acting on the probe tip can be calculated by,\cite{Iida2008ResonantForce}

\begin{equation}
  \mathbf{F}(\omega)
  = \frac{1}{2} \mathrm{Re}\!\Biggl[
    \sum_{i} \nabla \mathbf{E}(\mathbf{r}_{i}, \omega)^* \cdot \mathbf{P}(\mathbf{r}_{i}, \omega) V_i
  \Biggr]
  \tag{8}
\end{equation}

\noindent
where the summation was carried out over the cells assigned for probe metal tip.

\clearpage

\section*{Acknowledgements}

This work was supported by JSPS KAKENHI (Grant Numbers JP24K08282, JP24K21196, JP24K00933, JP25H01627, and JP26H00005).

\section*{Conflict of Interest}

The authors declare no conflict of interest.

\begin{shaded}
\noindent\textsf{\textbf{Keywords:} \keywords} 
\end{shaded}


\setlength{\bibsep}{0.0cm}
\bibliographystyle{Wiley-chemistry}
\bibliography{chemistry-template}

\begin{thebibliography}{10}

\bibitem{Verma2017}
P.~Verma, \emph{Chemical Reviews} \textbf{2017}, \emph{117}, 6447, pMID:
  28459149.

\bibitem{Lee2018}
H.~Lee, D.~Y. Lee, M.~G. Kang, Y.~Koo, T.~Kim, K.-D. Park, \emph{Nanophotonics}
  \textbf{2018}, \emph{9}, 3089.

\bibitem{Sifat2022}
A.~A. Sifat, J.~Jahng, E.~O. Potma, \emph{Chem. Soc. Rev.} \textbf{2022},
  \emph{51}, 4208.

\bibitem{Imada2016}
H.~Imada, K.~Miwa, M.~Imai-Imada, S.~Kawahara, K.~Kimura, Y.~Kim, \emph{Nature}
  \textbf{2016}, \emph{538}, 364.

\bibitem{Zhang2016}
Y.~Zhang, Y.~Luo, Y.~Zhang, Y.-J. Yu, Y.-M. Kuang, L.~Zhang, Q.-S. Meng,
  Y.~Luo, J.-L. Yang, Z.-C. Dong, J.~G. Hou, \emph{Nature} \textbf{2016},
  \emph{531}, 623.

\bibitem{Rajapaksa2010}
I.~Rajapaksa, K.~Uenal, H.~K. Wickramasinghe, \emph{Appl. Phys. Lett.}
  \textbf{2010}, \emph{97}, 073121.

\bibitem{Yamanishi2021}
J.~Yamanishi, H.~Yamane, Y.~Naitoh, Y.~Li, N.~Yokoshi, T.~Kameyama, S.~Koyama,
  T.~Torimoto, H.~Ishihara, Y.~Sugawara, \emph{Nature Communications}
  \textbf{2021}, \emph{12}, 3865.

\bibitem{Yamamoto2024}
T.~Yamamoto, H.~Yamane, N.~Yokoshi, H.~Oka, H.~Ishihara, Y.~Sugawara, \emph{ACS
  Nano} \textbf{2024}, \emph{18}, 1724.

\bibitem{Yamane2023}
H.~Yamane, N.~Yokoshi, H.~Oka, Y.~Sugawara, H.~Ishihara, \emph{Optics Express}
  \textbf{2023}, \emph{31}, 3415.

\bibitem{Yamanishi2023ChiroForce}
J.~Yamanishi, H.-Y. Ahn, H.~Okamoto, \emph{Nano Letters} \textbf{2023},
  \emph{23}, 9347.

\bibitem{Yamane2024SuperchiralPiFM}
H.~Yamane, M.~Hoshina, N.~Yokoshi, H.~Ishihara, \emph{The Journal of Chemical
  Physics} \textbf{2024}, \emph{160}, 044115, special Collection: Chirality of
  Plasmonic Structures and Materials.

\bibitem{Okamoto2019}
H.~Okamoto, \emph{Journal of Materials Chemistry C} \textbf{2019}, \emph{7},
  14771.

\bibitem{Purcell1973DDA}
E.~M. Purcell, C.~R. Pennypacker, \emph{The Astrophysical Journal}
  \textbf{1973}, \emph{186}, 705, provided by the SAO/NASA Astrophysics Data
  System.

\bibitem{Yurkin2007DDAReview}
M.~A. Yurkin, A.~G. Hoekstra, \emph{Journal of Quantitative Spectroscopy and
  Radiative Transfer} \textbf{2007}, \emph{106}, 558, iX Conference on
  Electromagnetic and Light Scattering by Non-Spherical Particles.

\bibitem{Chaumet2022DDAReview}
P.~C. Chaumet, \emph{Mathematics} \textbf{2022}, \emph{10}.

\bibitem{Tomoshige2025TEPL}
Y.~Tomoshige, M.~Tamura, H.~Ishihara, \emph{Applied Physics Express}
  \textbf{2025}, \emph{18}, 022004.

\bibitem{Tomoshige2025Nanophotonics}
Y.~Tomoshige, M.~Tamura, T.~Yokoyama, H.~Ishihara, \emph{Nanophotonics}
  \textbf{2025}, \emph{14}, 1157.

\bibitem{Ikagawa2026}
H.~Ikagawa, M.~Tamura, H.~Ishihara, \emph{Nano Letters} \textbf{2026},
  \emph{26}, 4613, pMID: 41837814.

\bibitem{Vial2008DCP}
A.~Vial, T.~Laroche, \emph{Applied Physics B} \textbf{2008}, \emph{93}, 139.

\bibitem{Johnson1972NobleMetals}
P.~B. Johnson, R.~W. Christy, \emph{Physical Review B} \textbf{1972}, \emph{6},
  4370.

\bibitem{Zhang2020}
Y.~Zhang, Z.-C. Dong, J.~Aizpurua, \emph{The Journal of Physical Chemistry C}
  \textbf{2020}, \emph{124}, 4674.

\bibitem{Lee2023}
M.-W. Lee, L.-Y. Hsu, \emph{Phys. Rev. A} \textbf{2023}, \emph{107}, 053709.

\bibitem{Schmidt1993GAMESS}
M.~W. Schmidt, K.~K. Baldridge, J.~A. Boatz, S.~T. Elbert, M.~S. Gordon, J.~H.
  Jensen, S.~Koseki, N.~Matsunaga, K.~A. Nguyen, S.~Su, T.~L. Windus,
  M.~Dupuis, J.~A. Montgomery~Jr., \emph{Journal of Computational Chemistry}
  \textbf{1993}, \emph{14}, 1347.

\bibitem{Iida2008ResonantForce}
T.~Iida, H.~Ishihara, \emph{Physical Review B} \textbf{2008}, \emph{77},
  245319.

\end{thebibliography}

\clearpage










\end{document}